\PassOptionsToPackage{unicode}{hyperref}
\PassOptionsToPackage{hyphens}{url}
\documentclass[
]{article}
\usepackage{amsmath,amssymb}
\usepackage{lmodern}
\usepackage{ifxetex,ifluatex}
\ifnum 0\ifxetex 1\fi\ifluatex 1\fi=0 
  \usepackage[T1]{fontenc}
  \usepackage[utf8]{inputenc}
  \usepackage{textcomp} 
\else 
  \usepackage{unicode-math}
  \defaultfontfeatures{Scale=MatchLowercase}
  \defaultfontfeatures[\rmfamily]{Ligatures=TeX,Scale=1}
\fi
\IfFileExists{upquote.sty}{\usepackage{upquote}}{}
\IfFileExists{microtype.sty}{
  \usepackage[]{microtype}
  \UseMicrotypeSet[protrusion]{basicmath} 
}{}
\makeatletter
\@ifundefined{KOMAClassName}{
  \IfFileExists{parskip.sty}{%
    \usepackage{parskip}
  }{
    \setlength{\parindent}{0pt}
    \setlength{\parskip}{6pt plus 2pt minus 1pt}}
}{
  \KOMAoptions{parskip=half}}
\makeatother
\usepackage{xcolor}
\IfFileExists{xurl.sty}{\usepackage{xurl}}{} 
\IfFileExists{bookmark.sty}{\usepackage{bookmark}}{\usepackage{hyperref}}
\hypersetup{
  pdftitle={Content-based subject classification at article level in biomedical context},
  pdfkeywords={open science, subject classification, fasttext, word
embeddings, fields of research},
  hidelinks,
  pdfcreator={LaTeX via pandoc}}
\usepackage{color}
\usepackage{fancyvrb}

\DefineVerbatimEnvironment{Highlighting}{Verbatim}{commandchars=\\\{\}}
\newenvironment{Shaded}{}{}

\newcommand{\DecValTok}[1]{\textcolor[rgb]{0.25,0.63,0.44}{#1}}

\newcommand{\FloatTok}[1]{\textcolor[rgb]{0.25,0.63,0.44}{#1}}

\newcommand{\NormalTok}[1]{#1}
\newcommand{\OperatorTok}[1]{\textcolor[rgb]{0.40,0.40,0.40}{#1}}

\newcommand{\StringTok}[1]{\textcolor[rgb]{0.25,0.44,0.63}{#1}}

\usepackage{longtable,booktabs,array}
\usepackage{calc} 
\usepackage{etoolbox}
\makeatletter
\patchcmd\longtable{\par}{\if@noskipsec\mbox{}\fi\par}{}{}
\makeatother
\IfFileExists{footnotehyper.sty}{\usepackage{footnotehyper}}{\usepackage{footnote}}
\makesavenoteenv{longtable}
\usepackage{graphicx}
\makeatletter
\def\maxwidth{\ifdim\Gin@nat@width>\linewidth\linewidth\else\Gin@nat@width\fi}
\def\maxheight{\ifdim\Gin@nat@height>\textheight\textheight\else\Gin@nat@height\fi}
\makeatother
\setkeys{Gin}{width=\maxwidth,height=\maxheight,keepaspectratio}
\makeatletter
\def\fps@figure{htbp}
\makeatother
\ifluatex
  \usepackage{selnolig}  
\fi
\newlength{\cslhangindent}
\newlength{\csllabelwidth}
 {
  \setlength{\parindent}{0pt}
  \ifodd #1 \everypar{\setlength{\hangindent}{\cslhangindent}}\ignorespaces\fi
  \ifnum #2 > 0
  \setlength{\parskip}{#2\baselineskip}
  \fi
 }%
 {}
\usepackage{calc}

\newenvironment{cslreferences}%
  {\setlength{\parindent}{0pt}%
  \everypar{\setlength{\hangindent}{\cslhangindent}}\ignorespaces}%
  {\par}

\title{Content-based subject classification at article level in
biomedical context}
\usepackage{authblk}
\author[%
  1%
  ]{%
  Eric Jeangirard%
}
\affil[1]{French Ministry of Higher Education, Research and Innovation,
Paris, France}
\date{April 2021}

\makeatletter
\def\@maketitle{%
  \newpage \null \vskip 2em
  \begin {center}%
    \let \footnote \thanks
         {\LARGE \@title \par}%
         \vskip 1.5em%
                {\large \lineskip .5em%
                  \begin {tabular}[t]{c}%
                    \@author
                  \end {tabular}\par}%
                                                \vskip 1em{\large \@date}%
  \end {center}%
  \par
  \vskip 1.5em}
\makeatother

\begin{document}
\maketitle
\begin{abstract}
Subject classification is an important task to analyze scholarly
publications. In general, mainly two kinds of approaches are used:
classification at a journal level and classification at the article
level. We propose a mixed approach, leveraging on embeddings technique
in NLP to train classifiers with article metadata (title, abstract,
keywords in particular) labelled with the journal-level classification
FoR (Fields of Research) and then apply these~classifiers at the article
level. We use this approach in the context of biomedical publications
using metadata from Pubmed. Fasttext classifiers are trained with FoR
codes and used to classify publications based on their available
metadata. Results show that using a stratification sampling strategy for
training help reduce the bias due to unbalanced field~distribution. An
implementation of the method is proposed on the repository
https://github.com/dataesr/scientific\_tagger
\end{abstract}


\hypertarget{introduction}{%
\section{1. Introduction}\label{introduction}}

Classifying scholarly literature in subjects or fields can have multiple
applications, for example helping for search and discovery in
bibliographic and bibliometric tools, creating indicators to understand
how a research area is structured, or detecting new trends. In our case,
we are building tools to support and help the steering of the Open
Science public policy in France. In the French Open Science Monitor
(Jeangirard 2019), a first attempt to classify publications is proposed
in order to monitor the open access trends at the discipline level. With
the COVID-19 crisis, the need to open up science in the health field has
been reaffirmed. A classification at a lower level in the biomedical
area is then needed to help to steer the Open Science public policy.

Several data providers already provide classifications, but one of the
rules we have adopted in this work (COSO 2018) is to apply open science
principles to monitor open science. Accordingly, we want to use and
re-release open data, and use and redistribute open code. In this sense,
using classifications from non-open and non-libre sources was not an
option.

Different approaches for subject classifications have been set up, both
at the journal level and the article level. At the journal level,
expert-based classification is possible, as the ARC journal list ((ARC)
2018) with university consultations and disciplinary groups feedback.
More algorithmic approaches, for example using journal-journal citations
pattern (Leydesdorff and Rafols 2009) have been implemented. Going to a
publication-level classification brings several advantages: first, it
should allow classifying every type of document, published in a
well-established journal or as a preprint. Also, it should identify more
accurately articles published in multidisciplinary journals like Science
or Nature (even though a single article can by itself be classified as
multidisciplinary). At the article level, clustering techniques are
often proposed, based on citation links like (Boyack et al. 2011) or
(Waltman and Eck 2012). More recently, content-based deep learning
techniques have been applied to Wikipedia articles (Semberecki and
Maciejewski 2017).

In this paper, we investigate an article-level classification method
based only on metadata like title, abstract and keywords (in the future,
we plan to do so using the full-text). This method is not depending on
citation indexes, and that could be, maybe, transposed to classify other
scholarly objects than scholarly publications.

\hypertarget{method}{%
\section{2. Method}\label{method}}

\hypertarget{data-sources}{%
\subsection{2.1 Data sources}\label{data-sources}}

We need a data source for the publications in order to extract their
metadata. For the moment, we have chosen a simple, reliable, and easy
source to harvest, which is Pubmed. Pubmed metadata can easily be
harvested through a public API, and the available metadata is rich,
including affiliations, abstracts, keywords, and MeSH (Medical Subject
Headings) (Medicine, n.d.).

There is also a need of a labelled database to train machine learning
algorithms. That is the tough part as there is no open, recent,
comprehensive, article-level tagged publications database. That is why
we have chosen to introduce a proxy to go in-between journal-level and
article-level tagging.

Through Excellence in Research for Australia (ERA), the Australian
Research Council (ARC) released the ERA 2018 Journal List ((Mercieca and
Macauley 2008), ((ARC) 2018)). This list associates, for more than
25,000 journals, up to 3 Field of Research (FoR) codes, or `MD' for
multidisciplinary journals. This journal-level information can be
transposed to an article-level database (all the articles with the same
ISSN will be assigned the same fields), which can be used to train
algorithms based on the other article-level metadata available, in
particular the title, abstract, keywords, and MeSH.

The Fields of Research (FoR) is a hierarchical classification, with
2-digits, 4-digits, and 6-digits classes. We use the 2-digits and
4-digits FoR codes from the ERA 2018 journal list. More than 150
4-digits FoR codes exist, and most of them are not relevant for a
health-specific classifier. ~

This source presents the merits of being quite recent, with a
hierarchical structure, and produced by a public organization under a
cc-by licence.

Selecting the FoR codes that are relevant for Pubmed papers (and
biomedical in general) is not an obvious task. Of course, we could limit
ourselves to FoR code 11 ``Medical and Health Sciences'' but that would
miss a lot of fields, in particular from Biology, Chemistry, and
Psychology. Asking for expert inputs could have been an option but it
seems there is no strong consensus on the perimeter. So we eventually
chose a quantitative approach, looking at the distribution of the FoR
codes in Pubmed data (based on the ISSN).

\hypertarget{training}{%
\subsection{2.2 Training}\label{training}}

Getting labelled data at the article level is very costly, because it
needs a high level of domain-specific expertise for a huge amount of
publications. At the same time, labelled data does exist at the journal
level. We try here to leverage that data to produce an article-level
classifier.

The idea is to extract relevant pieces of information~from available
metadata (title, abstract, keyword, MeSH) to classify a publication. We
assume that the publications of journals tagged with a given Field of
Research, ``Chemical Sciences'' for example, will contain field-specific
words and n-gram in their metadata that will be caught by a machine
learning classifier. As a consequence, if an article from, for example,
a ``Multidisciplinary'' journal contains enough words specific to
``Chemical Sciences'', then we guess that a machine learning approach
will be able to classify it as ``Chemical Sciences'' rather than
``Multidisciplinary''. In that case, the classification at the article
level would be different from the one at the journal level.

We propose to train multiple machine learning models, one for each
metadata type: title, ~abstract, keywords, MeSH, and journal title.

The design of the training dataset is key in the final model relevance
and performance. We evaluate two techniques to set up the training
dataset. First, with a simple random sampling. This way the distribution
of the classes in the training dataset is the same as in the whole data
set. In a second approach, with a stratification sampling, each class
being represented equally. The aim of this second sampling technique is
to lower the risk of bias: indeed, if the dataset is very unbalanced,
with one class being largely predominant, the risk the classifier
overfits this class can be high. Imagine an extreme case in which 99\%
of the cases are tagged with label A, and 1\% of the cases with label B.
Then a dummy classifier that always predicts A would have a 99\%
precision but would miss all the B cases. The idea of the stratified
sampling is to train the algorithm with 50\% of A cases and 50\% of B
cases to try to make the algorithm learn more relevant features.

Machine learning models are built with fasttext (Joulin et al. 2016)
from Facebook. It uses word embeddings techniques like in word2vec
models used in (Semberecki and Maciejewski 2017). Contrary to the latest
deep learning models, fasttext is extremely efficient on CPU and very
fast, and so is very adapted to low budgets environments.

\hypertarget{prediction}{%
\subsection{2.3 Prediction}\label{prediction}}

For the field prediction at the article level, we propose to combine the
outputs of each of the 5 models (for each metadata available) with a
voting system, giving slightly more weight to the journal title in case
of equality. Only the prediction with an associated score above a given
threshold is taken into account. As an example, if the model for title
and keywords predict ``Chemical Sciences'' (FoR 03) and the model based
on abstract and the one based on journal-title predict ``Biochemistry
and Cell Biology'' (FoR 0601), the heuristic to pick up the selected
field chooses the second one. But in another case, if the model for
journal-title predicts ``Multidisciplinary'' (MD) whereas models for
keywords, MeSH and title predict ``Psychology and Cognitive Sciences''
(FoR 17), then the heuristic returns the latter, giving a result
different from the pure journal-title information.

\hypertarget{results}{%
\section{3. Results}\label{results}}

\hypertarget{for-codes-selection}{%
\subsection{3.1 FoR codes selection}\label{for-codes-selection}}

After matching the ISSN from the Pubmed metadata and the ERA journal
list, we end up with an article-level database, enriched with Fields of
Research (matched at the ISSN level)

\newpage

\textbf{Table 1 : Sample of the available data}

\begin{longtable}[]{@{}llll@{}}
\toprule
\begin{minipage}[b]{0.22\columnwidth}\raggedright
pmid\strut
\end{minipage} & \begin{minipage}[b]{0.22\columnwidth}\raggedright
issn\strut
\end{minipage} & \begin{minipage}[b]{0.22\columnwidth}\raggedright
journal\_title\strut
\end{minipage} & \begin{minipage}[b]{0.22\columnwidth}\raggedright
FoR\strut
\end{minipage}\tabularnewline
\midrule
\endhead
\begin{minipage}[t]{0.22\columnwidth}\raggedright
31739602\strut
\end{minipage} & \begin{minipage}[t]{0.22\columnwidth}\raggedright
2076-2607\strut
\end{minipage} & \begin{minipage}[t]{0.22\columnwidth}\raggedright
Microorganisms\strut
\end{minipage} & \begin{minipage}[t]{0.22\columnwidth}\raggedright
-\strut
\end{minipage}\tabularnewline
\begin{minipage}[t]{0.22\columnwidth}\raggedright
31178264\strut
\end{minipage} & \begin{minipage}[t]{0.22\columnwidth}\raggedright
1950-6007\strut
\end{minipage} & \begin{minipage}[t]{0.22\columnwidth}\raggedright
Biomedicine \& pharmaco-therapy\strut
\end{minipage} & \begin{minipage}[t]{0.22\columnwidth}\raggedright
Pharmacology and Pharmaceutical Sciences\strut
\end{minipage}\tabularnewline
\begin{minipage}[t]{0.22\columnwidth}\raggedright
31218652\strut
\end{minipage} & \begin{minipage}[t]{0.22\columnwidth}\raggedright
1874-9356\strut
\end{minipage} & \begin{minipage}[t]{0.22\columnwidth}\raggedright
Folia microbiologica\strut
\end{minipage} & \begin{minipage}[t]{0.22\columnwidth}\raggedright
Microbiology ; Medical Microbiology\strut
\end{minipage}\tabularnewline
\begin{minipage}[t]{0.22\columnwidth}\raggedright
31669771\strut
\end{minipage} & \begin{minipage}[t]{0.22\columnwidth}\raggedright
1556-3871\strut
\end{minipage} & \begin{minipage}[t]{0.22\columnwidth}\raggedright
Heart rhythm\strut
\end{minipage} & \begin{minipage}[t]{0.22\columnwidth}\raggedright
Biomedical Engineering ; Cardiorespiratory Medicine and
Haematology\strut
\end{minipage}\tabularnewline
\begin{minipage}[t]{0.22\columnwidth}\raggedright
31473396\strut
\end{minipage} & \begin{minipage}[t]{0.22\columnwidth}\raggedright
1095-8630\strut
\end{minipage} & \begin{minipage}[t]{0.22\columnwidth}\raggedright
Journal of environmental management\strut
\end{minipage} & \begin{minipage}[t]{0.22\columnwidth}\raggedright
Multidisciplinary\strut
\end{minipage}\tabularnewline
\bottomrule
\end{longtable}

We use a sample of 500,000 records published in 2019 from the Pubmed
metadata to evaluate the following statistics.

First, it appears that 18\% of the publications in PubMed cannot be
attached directly to a FoR code: indeed their ISSN is not in ERA data,
so there is no correspondence between the ISSN and FoR codes.

\textbf{Table 2: Number of FoR matched to publications in Pubmed (sample
from 2019)}

\begin{longtable}[]{@{}ll@{}}
\toprule
Number of FoR & \% of publications\tabularnewline
\midrule
\endhead
No FoR & 18\%\tabularnewline
1 FoR & 37\%\tabularnewline
2 FoR & 23\%\tabularnewline
3 FoR & 22\%\tabularnewline
\bottomrule
\end{longtable}

At the 2-digits FoR code level, all the 150+ FoR codes are present in
the Pubmed data (from FoR 11 ``Medical and Health Sciences'' for 52\% of
the papers to FoR 19 ``Studies in Creative Arts and Writing'' for less
than 0.1\% of the papers). Note that a paper can be attached to multiple
FoR codes (up to 3).

\textbf{Table 3: Distribution of the 2-digits FoR codes in Pubmed
(sample from 2019)}

\begin{longtable}[]{@{}lll@{}}
\toprule
FoR & FoR code & Percentage of publications\tabularnewline
\midrule
\endhead
Medical and Health Sciences & 11 & 51.8\%\tabularnewline
Biological Sciences & 06 & 16.8\%\tabularnewline
Chemical Sciences & 03 & 11.3\%\tabularnewline
Engineering & 09 & 8.9\%\tabularnewline
Multidisciplinary & MD & 8.1\%\tabularnewline
Psychology and Cognitive Sciences & 17 & 7.2\%\tabularnewline
Physical Sciences & 02 & 3.2\%\tabularnewline
Agricultural and Veterinary Sciences & 07 & 3.1\%\tabularnewline
other & x & \textless{} 3\%\tabularnewline
\bottomrule
\end{longtable}

Concerning the~two main codes FoR 11 ``Medical and Health Sciences'' and
FoR 06 ``Biological Sciences'', we look deeper into the 4-digits FoR
codes. For the others, we selected only the FoR codes representing more
than 3\% of the papers in Pubmed. With the same logic, for the 4-digits
FoR codes in 11 - ``Medical and Health Sciences'' and 06 - ``Biological
Sciences'', we selected those representing more than 2\% of the papers,
and grouping the others into ``Other Medical and Health Sciences'' and
``Other Biological Sciences''.

After this selection process, we end up with the 17 fields presented in
Table 4 to classify publications in Pubmed.

\textbf{Table 4: Distribution of the selected fields in PubMed (sample
from 2019)}

\begin{longtable}[]{@{}lll@{}}
\toprule
Class & FoR code & Percentage of publications\tabularnewline
\midrule
\endhead
Clinical Sciences & 1103 & 19.9\%\tabularnewline
Chemical Sciences & 03 & 11.3\%\tabularnewline
Other Medical and Health Sciences & - & 10.3\%\tabularnewline
Engineering & 09 & 8.9\%\tabularnewline
Multidisciplinary & MD & 8.1\%\tabularnewline
Public Health and Health Services & 1117 & 7.4\%\tabularnewline
Psychology and Cognitive Sciences & 17 & 7.2\%\tabularnewline
Biochemistry and Cell Biology & 0601 & 5.4\%\tabularnewline
Neurosciences & 1109 & 4.5\%\tabularnewline
Other Biological Sciences & - & 4.5\%\tabularnewline
Pharmacology and Pharmaceutical Sciences & 1115 & 4.1\%\tabularnewline
Oncology and Carcinogenesis & 1112 & 3.7\%\tabularnewline
Cardiorespiratory Medicine and Haematology & 1102 & 3.3\%\tabularnewline
Physical Sciences & 02 & 3.2\%\tabularnewline
Agricultural and Veterinary Sciences & 07 & 3.1\%\tabularnewline
Paediatrics and Reproductive Medicine & 1114 & 2.6\%\tabularnewline
Microbiology & 0605 & 2.1\%\tabularnewline
\bottomrule
\end{longtable}

With this field selection, some publications do not get a field anymore
(the publications whose ISSN are not the selected scope). However, this
loss is very low (less than 1\%) as shown in Table 5.

\textbf{Table 5: Number of fields matched to publications in Pubmed}

\begin{longtable}[]{@{}lll@{}}
\toprule
Number of fields ~ & 2-digits FoR & Selected fields\tabularnewline
\midrule
\endhead
0 & 17.9\% & 18.7\%\tabularnewline
1 & 37.1\% & 56.6\%\tabularnewline
2 & 22.9\% & 20.9\%\tabularnewline
3 & 22.1\% & 3.8\%\tabularnewline
\bottomrule
\end{longtable}

\hypertarget{metadata-availability-in-pubmed}{%
\subsection{3.2 Metadata availability in
Pubmed}\label{metadata-availability-in-pubmed}}

In our approach, we assume rich metadata is available, in particular
metadata that can help infer a scientific discipline: title, abstract,
keywords, MeSH (and journal title). ~In the general case (on Crossref
for example), for most of the publications, the abstract and keywords
are not available. We look here at Pubmed metadata.

\textbf{Table 6: Metadata availability in Pubmed (sample from 2019
publications)}

\begin{longtable}[]{@{}ll@{}}
\toprule
Metadata & Availability\tabularnewline
\midrule
\endhead
Abstract ~ & 87.5\%\tabularnewline
Keywords & 64.2\%\tabularnewline
MeSH & 69.1\%\tabularnewline
at least 1 among abstract, keywords, MeSH & 96.6\%\tabularnewline
at least 2 among abstract, keywords, MeSH & 80.9\%\tabularnewline
all 3 metadata & 43.3\%\tabularnewline
\bottomrule
\end{longtable}

So the coverage is far from perfect, but in more than 80\%, 2 out of 3
key metadata are available.

Field-wise, there is of course a variety of situations as shown in
Figure 2. However, most of the selected fields have a good metadata
availability, except from ``Physical Sciences'' (FoR 02) with less than
30\% of publications with keywords available (but abstract is available
in more than 95\%). As a consequence, the situation is not perfect
(100\% availability would be a better scenario) but rich metadata (at
least partially) remain available in the vast majority of the cases.

\begin{figure}
\centering
\includegraphics[width=4.16667in,height=\textheight]{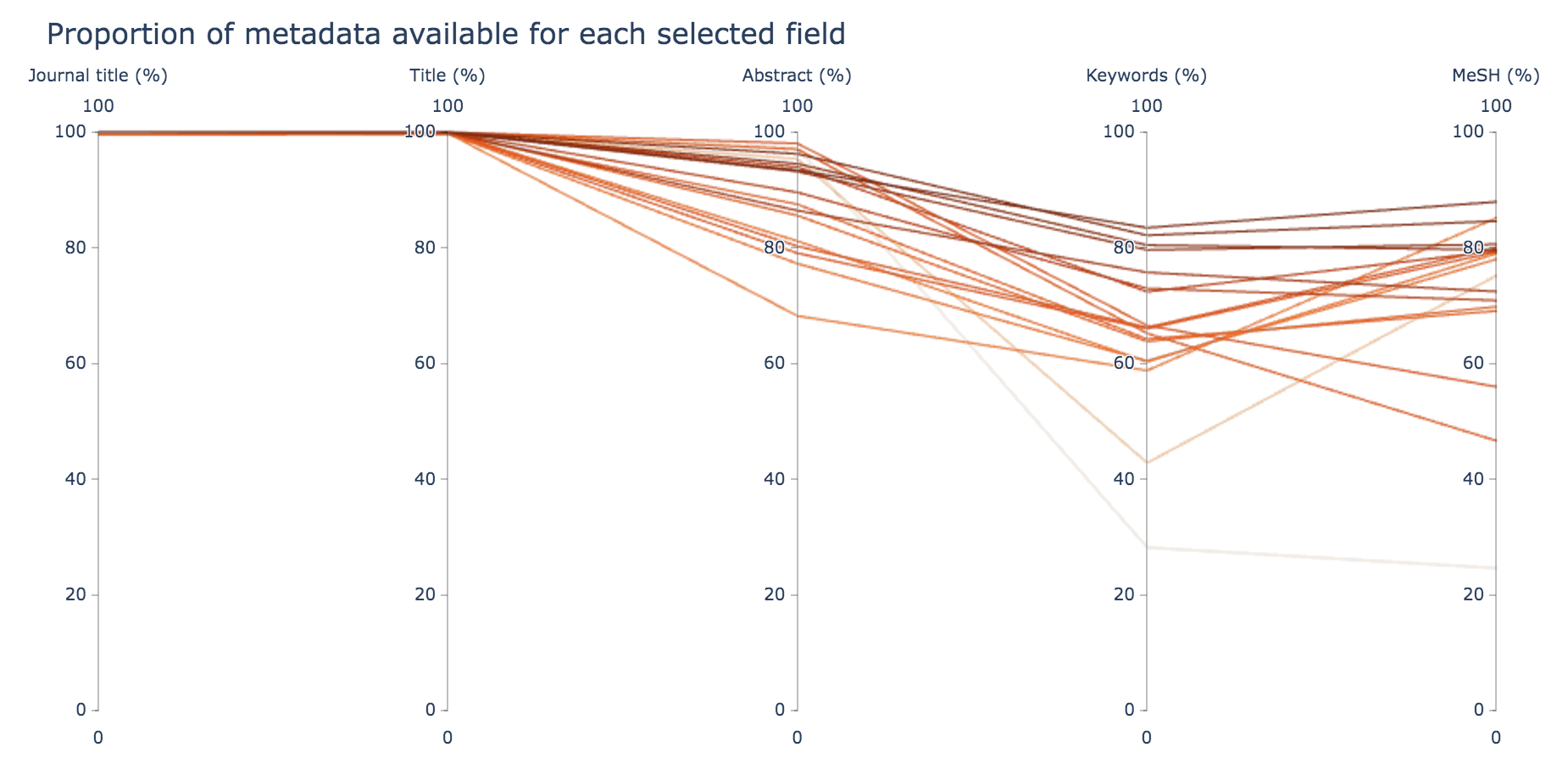}
\caption{Proportion of publications with metadata available in Pubmed
for the selected fields}
\end{figure}

\hypertarget{training-data}{%
\subsection{3.3 Training data}\label{training-data}}

We used two training datasets of each 850,000 publications. Each of them
is split for training and testing (90\%, 10\%). The first dataset is a
random sample from the Pubmed metadata (with at least a selected field
assigned). The second dataset is a stratified sample, each selected
field representing an equal part of the total.

For each dataset, we trained 5 \emph{fasttext} models, for each metadata
type: title, abstract, keywords, MeSH, and journal-title. The training
parameters are presented in Table 7, the other being set to default
values. All the possible parameters and their effects are detailed in
fasttext online documentation (Facebook, n.d.).

\textbf{Table 7: fasttext parameters used for training}

\begin{longtable}[]{@{}ll@{}}
\toprule
Parameter & Value\tabularnewline
\midrule
\endhead
epoch & 50\tabularnewline
wordNgrams & 2\tabularnewline
loss & ova\tabularnewline
minCount & 20\tabularnewline
\bottomrule
\end{longtable}

Several metrics can measure the performance of a classifier. In
particular, the precision computes the fraction of predicted classes
that are relevant and the recall that computes the fraction of relevant
classes that are successfully predicted. The f1 score combines precision
and recall ~(it is the harmonic mean of precision and recall). We report
the f1 score of each model in Table 8.

\newpage

\textbf{Table 9: f1 score of trained model}

\begin{longtable}[]{@{}llllll@{}}
\toprule
\begin{minipage}[b]{0.14\columnwidth}\raggedright
Sampling technique\strut
\end{minipage} & \begin{minipage}[b]{0.14\columnwidth}\raggedright
Journal title model\strut
\end{minipage} & \begin{minipage}[b]{0.14\columnwidth}\raggedright
Title model\strut
\end{minipage} & \begin{minipage}[b]{0.14\columnwidth}\raggedright
Abstract model\strut
\end{minipage} & \begin{minipage}[b]{0.14\columnwidth}\raggedright
Keywords model\strut
\end{minipage} & \begin{minipage}[b]{0.14\columnwidth}\raggedright
MeSH model\strut
\end{minipage}\tabularnewline
\midrule
\endhead
\begin{minipage}[t]{0.14\columnwidth}\raggedright
Random ~\strut
\end{minipage} & \begin{minipage}[t]{0.14\columnwidth}\raggedright
99.7\strut
\end{minipage} & \begin{minipage}[t]{0.14\columnwidth}\raggedright
40.8\strut
\end{minipage} & \begin{minipage}[t]{0.14\columnwidth}\raggedright
44.6\strut
\end{minipage} & \begin{minipage}[t]{0.14\columnwidth}\raggedright
43.6\strut
\end{minipage} & \begin{minipage}[t]{0.14\columnwidth}\raggedright
42.8\strut
\end{minipage}\tabularnewline
\begin{minipage}[t]{0.14\columnwidth}\raggedright
Stratified\strut
\end{minipage} & \begin{minipage}[t]{0.14\columnwidth}\raggedright
99.9\strut
\end{minipage} & \begin{minipage}[t]{0.14\columnwidth}\raggedright
52.4\strut
\end{minipage} & \begin{minipage}[t]{0.14\columnwidth}\raggedright
56.3\strut
\end{minipage} & \begin{minipage}[t]{0.14\columnwidth}\raggedright
53.6\strut
\end{minipage} & \begin{minipage}[t]{0.14\columnwidth}\raggedright
52.5\strut
\end{minipage}\tabularnewline
\bottomrule
\end{longtable}

We observe that the stratified sampling approach gives overall better
performance on each model.

We also notice that models based on journal-title have almost a perfect
f1 score. That can be explained as the model simply tries to replicate a
simple correspondence between ISSN (and so journal titles) and assigned
FoR. The only advantage of the machine learning approach rather than the
simple ISSN - FoR correspondence is the generalization for out-of-sample
journals. The model is still able to predict a field, even for journals
that are not part of the ERA 2018 journal list.

Apart from the journal-title model, the performance for the other model
could seem low. Actually, we have to keep in my mind that a perfect f1
score is not even the objective, as, in that hypothetic case, the
prediction would be exactly the same as the one based only on
journal-title.

\hypertarget{gaining-insights-on-calibrated-fasttext-embeddings}{%
\subsection{3.4 Gaining insights on calibrated fasttext
embeddings}\label{gaining-insights-on-calibrated-fasttext-embeddings}}

Fasttext is a word embeddings model, meaning that each word is
represented by a numeric vector, in our case in dimension 100. The
strength of this type of model is that the numeric distance between
these vectors can be interpreted as a semantic distance.

The fasttext library comes with two handy functions to explore the word
embeddings: \emph{get\_nearest\_neighbors} and \emph{get\_analogies}.
The first function lists the words whose embeddings are the closest to
the input. For instance, using the model calibrated on titles, the 3
nearest neighbors of ``infants'' are ``pregnancies'', ``childhood'' and
``children''.

\begin{Shaded}
\begin{Highlighting}[]
\NormalTok{model.get\_nearest\_neighbors(}\StringTok{"infants"}\NormalTok{)[}\DecValTok{0}\NormalTok{:}\DecValTok{3}\NormalTok{]}
\end{Highlighting}
\end{Shaded}

\begin{Shaded}
\begin{Highlighting}[]
\NormalTok{[(}\FloatTok{0.9249277710914612}\OperatorTok{,} \StringTok{\textquotesingle{}pregnancies\textquotesingle{}}\NormalTok{)}\OperatorTok{,}
\NormalTok{ (}\FloatTok{0.912368893623352}\OperatorTok{,} \StringTok{\textquotesingle{}children\textquotesingle{}}\NormalTok{)}\OperatorTok{,}
\NormalTok{ (}\FloatTok{0.9030213356018066}\OperatorTok{,} \StringTok{\textquotesingle{}childhood\textquotesingle{}}\NormalTok{)]}
\end{Highlighting}
\end{Shaded}

The other function enables the user to play around with word analogies.
Fasttext documentation (Facebook, n.d.) gives the example of the triplet
(``berlin'', ``germany'', ``france''), which can be interpreted as:
``What is to France what Berlin is to Germany ?''. In fasttext
documentation, the first result given by the model they use is
``paris''. We played the same game with the model calibrated on titles,
with the triplet (``hypertension'', ``heart'', ``brain''). That is to
say, according to the model we calibrated, what is to the brain what
hypertension is for the heart? The two first results are ``stroke'' and
``aneurysms''.

\begin{Shaded}
\begin{Highlighting}[]
\NormalTok{model.get\_analogies(}\StringTok{"hypertension"}\NormalTok{, }\StringTok{"heart"}\NormalTok{, }\StringTok{"brain"}\NormalTok{)[}\DecValTok{0}\NormalTok{:}\DecValTok{3}\NormalTok{]}
\end{Highlighting}
\end{Shaded}

\begin{Shaded}
\begin{Highlighting}[]
\NormalTok{[(}\FloatTok{0.9254266619682312}\OperatorTok{,} \StringTok{\textquotesingle{}stroke\textquotesingle{}}\NormalTok{)}\OperatorTok{,}
\NormalTok{ (}\FloatTok{0.913159966468811}\OperatorTok{,} \StringTok{\textquotesingle{}cerebral\textquotesingle{}}\NormalTok{)}\OperatorTok{,}
\NormalTok{ (}\FloatTok{0.9120014905929565}\OperatorTok{,} \StringTok{\textquotesingle{}aneurysms\textquotesingle{}}\NormalTok{)]}
\end{Highlighting}
\end{Shaded}

\hypertarget{classification-inference}{%
\subsection{3.5 Classification
inference}\label{classification-inference}}

We applied the classification method on 45,000 publications from Pubmed
with a French affiliation. For each one, we computed the field inferred
with only the journal title and the field predicted using the
combination of the 5 models (one for each metadata). For each model, we
keep only the predictions with a probability above 0.5. The probability
of each tag is directly computed by the fasttext library. We then look
at the transition matrix between journal-based and article-based
classification.

\begin{figure}
\centering
\includegraphics[width=4.16667in,height=\textheight]{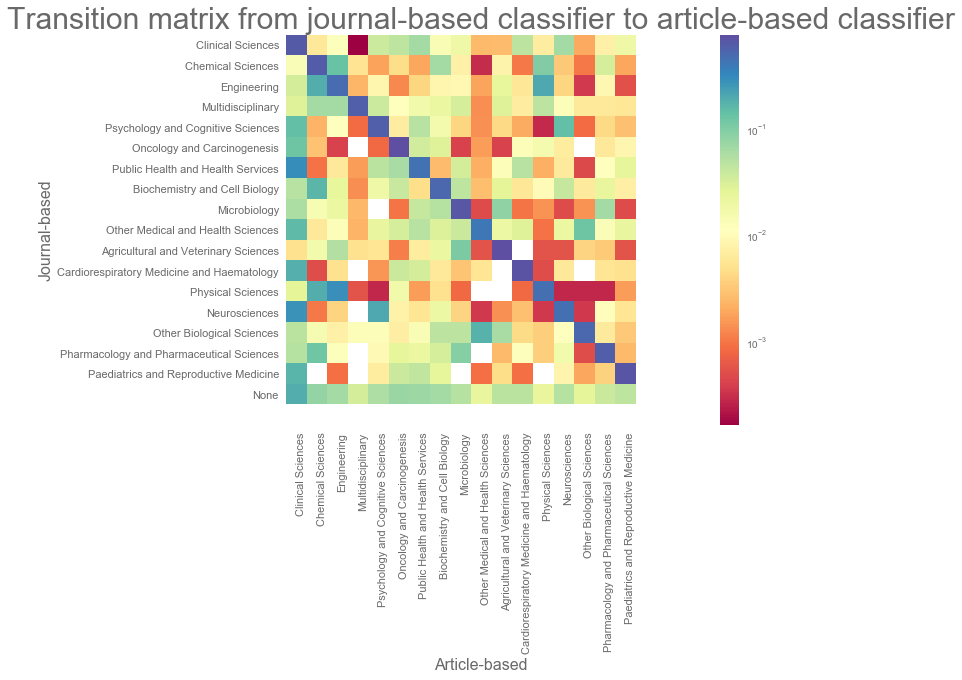}
\caption{Transition matrix from journal-based to article-based
classifier}
\end{figure}

As expected, in the majority of the cases, the article-based prediction
is the same as the journal level one. However, a few things can be
noticed. All the articles with no prediction at the journal level are
classified with the article-level approach. The likelihood of a label
change between the journal level and the article level is higher for
some fields than others (in particular, ~``Multidisciplinary'' or
``Other Medical and Health Services''). For example, a publication whose
title is `Topography and behavioral relevance of the global signal in
the human brain.', published in `Scientific reports' was classified
``Multidisciplinary'' with a journal-based model and became classified
``Psychology and Cognitive Sciences'' with the combination of the
models. Also, some transitions are more frequent, like ``Physical
Sciences'' to ``Engineering'' or ``Neurosciences'' to ``Psychology and
Cognitive Sciences''.

\hypertarget{discussion-and-conclusion}{%
\section{4. Discussion and conclusion}\label{discussion-and-conclusion}}

In this paper, we show how we combined existing and open data sources
(Pubmed, Fields of Research) to train a machine learning model able to
classify publications in the area of medicine and health. We proposed a
mixed approach between journal-based and article-based classification
and apply it to the metadata of French publications in Pubmed. This work
was done as a pre-requisite for the construction of the French Open
Science Monitor in Health and Medicine that will be released later. We
propose a python implementation in
https://github.com/dataesr/scientific\_tagger

\hypertarget{findings}{%
\subsection{4.1 Findings}\label{findings}}

We first have shown that a combination of 17 Field of Research (2-digits
and 4-digits) have good coverage of the publications in Pubmed, and
propose to use these 17 fields for building a biomedical publication
classifier.

We also underlined the role of the construction of the training dataset
and put in place a technique to handle unbalanced class distribution
using stratification.

With a few examples, we show the ability of fasttext models to manage to
encode in their word embeddings at least part of the meaning of the
words used in publications metadata (in the title in particular).

Finally, we have constructed a heuristic to predict a subject at the
article level, using pieces of information~from the available metadata
(title, abstract, keywords, MeSH, and journal title). This method is
most of the time in line~with a pure journal-based classification, but
brings two main advantages, being able to predict a class even if the
journal is unknown and handling a bit better the articles published in
Multidisciplinary journals.

\hypertarget{limitations-and-future-research}{%
\subsection{4.2 Limitations and future
research}\label{limitations-and-future-research}}

The main limit of this work is the difficulty to evaluate the relevance
of the final classification. Indeed, only a manual check on a
representative sample of several hundreds of publications would allow to
really measure the validity of the approach. That is part of future work
we would like to conduct, like (Bornmann 2018) but on a larger scale to
get insights on the potential bias of our method.

Another way to test the relevance of the output classes would be to use
citation patterns to confirm or infirm the class's predictions like in
(Wang and Waltman 2016).

Finally, we have to keep in mind that at the time of writing of (Joulin
et al. 2016), fasttext was very efficient and on par with more complex
deep learning architecture, but this area is evolving extremely fast and
more recent NLP techniques should be investigated as well.

\hypertarget{software-and-code-availability}{%
\section{Software and code
availability}\label{software-and-code-availability}}

The source code is released under an MIT license in the GitHub
repository https://github.com/dataesr/scientific\_tagger

\hypertarget{references}{%
\section*{References}\label{references}}
\addcontentsline{toc}{section}{References}

\hypertarget{refs}{}
\begin{cslreferences}
\leavevmode\hypertarget{ref-australian_research_council_arc_era_2018}{}%
(ARC), Australian Research Council. 2018. ``ERA 2018 Journal List.''
\url{https://www.arc.gov.au/excellence-research-australia/era-2018-journal-list}.

\leavevmode\hypertarget{ref-bornmann_field_2018}{}%
Bornmann, Lutz. 2018. ``Field Classification of Publications in
Dimensions: A First Case Study Testing Its Reliability and Validity.''
\emph{Scientometrics} 117 (1): 637--40.
\url{https://doi.org/10.1007/s11192-018-2855-y}.

\leavevmode\hypertarget{ref-boyack_clustering_2011}{}%
Boyack, Kevin W., David Newman, Russell J. Duhon, Richard Klavans,
Michael Patek, Joseph R. Biberstine, Bob Schijvenaars, André Skupin,
Nianli Ma, and Katy Börner. 2011. ``Clustering More Than Two Million
Biomedical Publications: Comparing the Accuracies of Nine Text-Based
Similarity Approaches.'' Edited by Colin Allen. \emph{PLoS ONE} 6 (3):
e18029. \url{https://doi.org/10.1371/journal.pone.0018029}.

\leavevmode\hypertarget{ref-coso_feedback_2018}{}%
COSO, French Open Science Committee. 2018. ``Feedback on EC Open Science
Monitor Methodological Note.''
\url{https://www.ouvrirlascience.fr/feedback-ec-science-monitor/}.

\leavevmode\hypertarget{ref-facebook_fasttext_nodate}{}%
Facebook. n.d. ``Fasttext Documentation.''
\url{https://fasttext.cc/docs/en/supervised-tutorial.html}.

\leavevmode\hypertarget{ref-jeangirard_monitoring_2019}{}%
Jeangirard, Eric. 2019. ``Monitoring Open Access at a National Level:
French Case Study.'' In \emph{ELPUB 2019 23d International Conference on
Electronic Publishing}. OpenEdition Press.
\url{https://doi.org/10.4000/proceedings.elpub.2019.20}.

\leavevmode\hypertarget{ref-joulin_bag_2016}{}%
Joulin, Armand, Edouard Grave, Piotr Bojanowski, and Tomas Mikolov.
2016. ``Bag of Tricks for Efficient Text Classification.''
\emph{arXiv:1607.01759 {[}Cs{]}}, August.
\url{http://arxiv.org/abs/1607.01759}.

\leavevmode\hypertarget{ref-leydesdorff_global_2009}{}%
Leydesdorff, Loet, and Ismael Rafols. 2009. ``A Global Map of Science
Based on the ISI Subject Categories.'' \emph{Journal of the American
Society for Information Science and Technology} 60 (2): 348--62.
\url{https://doi.org/10.1002/asi.20967}.

\leavevmode\hypertarget{ref-us_nation_library_of_medicine_medical_nodate}{}%
Medicine, U. S. Nation Library of. n.d. ``Medical Subject Heading
(MeSH).'' \url{https://www.nlm.nih.gov/pubs/factsheets/mesh.html}.

\leavevmode\hypertarget{ref-mercieca_new_2008}{}%
Mercieca, Paul, and Peter Macauley. 2008. ``A New Era of Open Access?''
\emph{Australian Academic \& Research Libraries} 39 (4): 243--52.
\url{https://doi.org/10.1080/00048623.2008.10721362}.

\leavevmode\hypertarget{ref-semberecki_deep_2017}{}%
Semberecki, Piotr, and Henryk Maciejewski. 2017. ``Deep Learning Methods
for Subject Text Classification of Articles.'' \emph{Proceedings of the
2017 Federated Conference on Computer Science and Information Systems},
ACSIS, 11: 357--60. \url{https://doi.org/10.15439/2017F414}.

\leavevmode\hypertarget{ref-waltman_new_2012}{}%
Waltman, Ludo, and Nees Jan van Eck. 2012. ``A New Methodology for
Constructing a Publication-Level Classification System of Science: A New
Methodology for Constructing a Publication-Level Classification System
of Science.'' \emph{Journal of the American Society for Information
Science and Technology} 63 (12): 2378--92.
\url{https://doi.org/10.1002/asi.22748}.

\leavevmode\hypertarget{ref-wang_large-scale_2016}{}%
Wang, Qi, and Ludo Waltman. 2016. ``Large-Scale Analysis of the Accuracy
of the Journal Classification Systems of Web of Science and Scopus.''
\emph{Journal of Informetrics} 10 (2): 347--64.
\url{https://doi.org/10.1016/j.joi.2016.02.003}.
\end{cslreferences}

\end{document}